\documentclass[prb,superscriptaddress,twocolumn,showpacs]{revtex4}
\usepackage{amssymb}
\usepackage {longtable}
\usepackage{amsmath}
\usepackage{graphics}
\usepackage{graphicx}
\usepackage{dcolumn}

\begin{document}
\title{Relaxation of high-energy quasiparticle distributions:
electron-electron scattering in a two-dimensional electron gas}

\author{R.N.Gurzhi}
\affiliation{B.Verkin Institute for Low Temperature Physics \&
Engineering,  National Academy of Sciences of Ukraine, 47 Lenin
Ave, Kharkov, 61103, Ukraine}
\author{A.I.Kopeliovich}
\affiliation{B.Verkin Institute for Low Temperature Physics \&
Engineering, National Academy of Sciences of Ukraine, 47 Lenin
Ave, Kharkov, 61103, Ukraine}
\author{A.N.Kalinenko}
\affiliation{B.Verkin Institute for Low Temperature Physics \&
Engineering,  National Academy of Sciences of Ukraine, 47 Lenin
Ave, Kharkov, 61103, Ukraine}
\author{A.V.Yanovsky}
\affiliation{B.Verkin Institute for Low Temperature Physics \&
Engineering,  National Academy of Sciences of Ukraine, 47 Lenin
Ave, Kharkov, 61103, Ukraine}
\author{E.N.Bogachek}
\affiliation{School of Physics, Georgia Institute of Technology,
Atlanta, GA 30332-0430, USA}
\author{Uzi Landman}
\affiliation{School of Physics, Georgia Institute of Technology,
Atlanta, GA 30332-0430, USA}
\author{H. Buhmann}
\affiliation{Physikalisches Institut der Universit\"at W\"urzburg,
D-97047 W\"urzburg, Germany}
\author{L.W. Molenkamp}
\affiliation{Physikalisches Institut der Universit\"at W\"urzburg,
D-97047 W\"urzburg, Germany}

\begin{abstract}
A theory is developed  for the evolution of the non-equilibrium
 distribution of quasiparticles  when the scattering
rate decreases due to particle collisions. We propose a  "modified
one-collision approximation" which is most effective for
high-energy quasiparticle distributions. This method is used to
explain novel measurements of the non-monotonic energy dependence
of the signal of scattered electrons  in a 2D system. The observed
effect is related to a crossover from the ballistic to the
hydrodynamic regime of electron flow.
\end{abstract}

 \maketitle
\newpage
\section{Introduction}

The propagation of an electron beam in a two-dimensional
degenerate electron gas (2DEG) can be used to study the role of
electron-electron (e-e) collisions in reduced dimensions, see e.g.
Refs. \cite{msst,juapp,cumming}. Of main interest are
"high-energy" beams with a typical electron energy that exceeds
considerably the thermal energy of the equilibrium electrons
within the 2DEG, but remains significantly below the Fermi energy.
A special combination of ac- and dc-measurement techniques allows
for an energy selective investigation of electrons traversing a
defined distance within a 2DEG \cite{epl2001}. Some of these
experiments were analyzed in terms of the $\tau$-approximation
\cite{chaplik,quinn}, where it is supposed that each collision
scatters an electron out of the beam. However, this approach is
obviously insufficient for the case of electron-electron (ee)
scattering in a 2DEG, where small angle scattering events
dominate, and scattered electrons may remain within the region of
the electron beam \cite{epl2001,gfnt97,gprl,gprb2000,gphyse98}.

For an adequate theoretical description of ee-scattering it is
necessary to take into account the incoming terms in the collision
operator $\hat {J}$ (see the second term on the right of the
equation for $\hat {J}$)  of the dynamic integro-differential
linearized Boltzmann equation:

\begin{eqnarray}\label{e-kin}
    \frac{d f}{d t} = \hat{J} f , & & \nonumber\\
    \hat{J}f = -\nu f + \int d{\bf p'}
    \nu_{\bf p p'}f({\bf p'}), \ \nu = \int d{\bf p'} \nu_{\bf p' p} \
    ,  \\   \frac{d f}{d t} \equiv  \frac{\partial f}{\partial t}+
    {\bf v}\frac{\partial f}{\partial {\bf r}} + {\bf F}
    \frac{\partial f}{\partial {\bf p}} & &\nonumber.
\end{eqnarray}

Here, $f$ is the distribution function of non-equilibrium
electrons, ${\bf r}$ determines the location of a non-equilibrium
electron, and ${\bf v}$ and ${\bf p}$ are the electron velocity
and momentum, respectively (quasi-classical approximation). ${\bf
F}$ is the Lorentz force acting on the electron. The function
$\nu_{\bf p' p}$ determines the probability of non-equilibrium
electrons ($\nu_{\bf p' p}>0$) or holes ($\nu_{\bf p'p}<0$) to be
found in state ${\bf p}$ after scattering from state ${\bf p'}$. A
hole in this case describes the absence of an equilibrium electron
below the Fermi surface.

\begin{figure}
\includegraphics[height=6 cm,width=8cm]{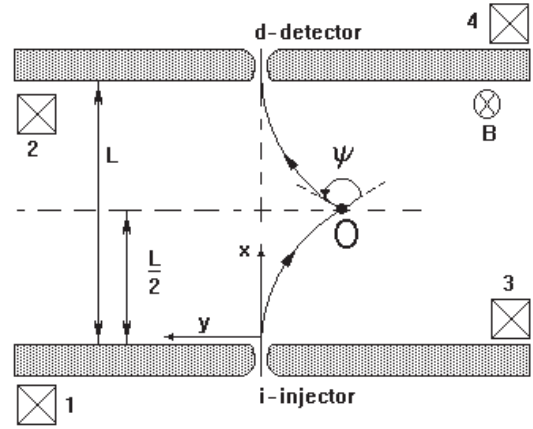}
\caption{Scheme of the e-e spectroscopy experiment (see text).}
\label{g-exp}
\end{figure}

In the general case, the solution of the kinetic Boltzmann
equation (\ref{e-kin}) is a very difficult task, both in
analytical and numerical calculations. But it is well known from
the kinetics of gases that two regimes exist which allow for an
essential simplification of the description. First, in the
few-collision case the probability is small for an electron to be
scattered over a characteristic length scale $L$ ( e.g., the
distance between the injector to the detector, see Fig.\ref{g-exp}
below), which is much smaller than the electron-electron mean-free
path, $l$, i.e.\ $l = v \nu ^ {-1} \gg L $. Therefore, using
perturbation theory it is possible to solve the Boltzmann equation
by iteration of the collision operator (quasiballistic
approximation). Second, in the case of frequent collisions  ($l
\ll L $) a hydrodynamic situation arises. The influence of
frequent collisions establishes a quasi-equilibrium distribution
which is characterized by hydrodynamic variables, such as the
average velocity or temperature which depend only smoothly on
${\bf r}$. These regimes were studied in our previous publications
\cite{epl2001,gprl,gphyse98,gprb2000}. For obvious reasons a
treatment of the intermediate regime ($l\approx L$) presents
considerable difficulties.

In this paper we show that it is possible to study the
intermediate regime successfully if the scattering probability
decreases substantially with each collision. The occurrence of a
decreasing scattering probability can be expressed by an
additional small parameter $ \Delta $ which is defined as the
ratio of the scattering probability of a nonequilibrium electron
after a collision to the probability before this event. Such
situations can be found in a wide range of problems where
non-equilibrium quasi-particles exhibiting some excess energy,
above thermal equilibrium, collide with equilibrium
quasi-particles and lose a considerable amount of this energy due
to this event. Usually, the probability for inelastic scattering
decreases drastically with decreasing energy of the
quasi-particles, because the statistical weight decreases for both
the possibility for finding a collision partner and for finding
empty final states. Electron-electron, electron-phonon or
phonon-phonon collisions at temperatures below the Debye
temperature can be taken as an example.

Our approach is based on the following concept: (i) a few
collisions are treated exactly, while (ii) subsequent collisions
are neglected, or considered perturbationally. We will refer to
this approach as the "modified one-collision approximation" (MOC),
which accounts for the fact that in most cases a restriction to an
one collision approximation is sufficient to explain the general
features of the scattering experiments. In the present work, a
full account of the MOC approximation is given and the results are
applied to analyze experimental data obtained from electron beam
experiments. It is shown, that applicability of the MOC
approximation to the scattering of high-energy electrons
substantially extends the description of spectroscopic data from
electron-electron collisions experiments \cite {epl2001}. The
non-monotonic dependence of the signal of scattered electrons on
the beam energy can be explained within the MOC approximation.
This observation can be interpreted as a crossover from the
ballistic to multi-collision (hydrodynamic) flow regime.

\section{Modified One-Collision Approximation}

Let us consider non-equilibrium electrons at a characteristic
energy $\varepsilon_{0}$, measured from the Fermi level
$\varepsilon_{F}$, and larger than the temperature $T$ of the
equilibrium electron gas, $\varepsilon_0 \gg T$. In a first stage
of relaxation these highly non-equilibrium electrons are scattered
by equilibrium electrons thus generating a new non-equilibrium
system but with already much lower excess energy. This stage of
relaxation was named the "pretemperature" stage in
Ref.\cite{gfnt97}. This "pretemperature" stage ends when the
quasi-particles energy becomes comparable to the bath temperature
$T$. The most important fact for the "pretemperature" stage is the
following: If the energy $\varepsilon_{0} < \varepsilon_{F}$, then
the probability of secondary collision is an order of magnitude
lower than for the first one \cite{gfnt97}. Indeed, the energy
$\varepsilon$ of a non-equilibrium electron after collision with
an equilibrium electron will be distributed between three states
in phase space:

 $$ {\bf p} + {\bf p'} \rightarrow {\bf p''} + {\bf p'''} \ ;$$
 $$ \varepsilon = |\varepsilon'| + \varepsilon'' + \varepsilon''' \ . $$

For $T\ll \varepsilon$, we have $\varepsilon'< 0$ and
$\varepsilon'',\varepsilon''' > 0$ due to Pauli principle, i.e.
the state ${\bf p'}$ is below the Fermi level while the final
states ${\bf p''}$ and ${\bf p}'''$ are well above. The energy of
the initial electron is redistributed  between three final states.
On average, the energy of each state after the first collision can
be expressed as $\varepsilon_1 \sim \varepsilon_0 \cdot \Delta$,
with $\Delta^2 \approx 10^{-1}$ (Note, the numerical value of
$\Delta$ depends on the specific type of problem
considered)\cite{gfnt97}. Since the energy dependent mean-free
path $l(\varepsilon) \propto 1/\varepsilon^2$, this means that
$l(\varepsilon)$ of non-equilibrium quasi-particles will increase
by about an order of magnitude, which implies that
$l(\varepsilon_1) \approx \Delta^{-2} \cdot l ( \varepsilon_0 )\gg
l(\varepsilon_0)$. This may lead to a situation where the
probability for a secondary collision is low while the probability
for the first collision is close to unity in a system of size $L$:
$l(\varepsilon_1) \gg L \gtrsim l(\varepsilon_0)$. In this case,
it is a natural choice to consider the first collision exactly
while the secondary collision is treated by perturbation theory,
which leads to an extension of the validity of the one-collision
approximation up to $L \sim 10 \ l(\varepsilon_0)$.

In the following, we use this approach for the case of propagation
of nonequilibrium electrons in an electron beam injected into a
2DEG via a quantum point-contact. For steady state conditions, the
operator $d/dt$ on the left side of Eq.~(\ref {e-kin}) must be
replaced by $({\bf v
\partial/\partial r}) + ({\bf F \partial/\partial p})$. For
further consideration it is convenient to use Cartesian
coordinates. The x-axis is along the beam direction, while the
origin is located at the center of the electron beam injector (a
quantum point-contact in an actual experiment). Eq.~(\ref{e-kin})
will be solved by integrating along the trajectory. We define $t$
as the time of motion of the particle along a given trajectory.
${\bf R}({\bf r},{\bf p}, t)$ is the (Lagrange) coordinate of the
particle at a time $t$. At the time of observation (taken as the
origin of the time counting), the particle is located at the point
${\bf r}$ with a momentum ${\bf p}$. The momentum of the particle
at a time $t$ is ${\bf P}({\bf r}, {\bf p}, t) = m {\bf V} =
m(\partial {\bf R}/\partial t)$. The particle trajectory ${\bf
R}({\bf r},{\bf p}, t)$ is uniquely defined for given applied
fields. The solution of the equation $X({\bf r}, {\bf p}, t_0)=0$
defines the start time $t_0$ of the electron injection into the
system at $x=0$. The solution of Eq.~(\ref{e-kin}) is given as

\begin{equation*}
 \label{e-mocriad}
f = \sum_{n=0}^{\infty}f^{(n)},
\end{equation*}
\begin{gather}
 f^{(n+1)}({\bf r},
{\bf p}) = \int_{t_0}^0 d \ t \int d{\bf p'} [\nu_{\bf P p'} -
\delta({\bf P} -{\bf p'}) \ \nu ] f^{(n)}({\bf R}, {\bf p'})\nonumber\\
\equiv
 \int_{t_0}^0 dt (\hat{\nu} - \nu) f^{(n)}.
\end{gather}

In this expression each $n$-th term of the series is obtained by
$n$ subsequent actions of the collision operator on the initial or
unpertubated function $f^{(0)}$ which describes the ballistic
motion of the particle:

\begin{eqnarray}\label{e-balsolv}
f^{(0)}({\bf r}, {\bf p}) = f_0(Y_0,{\bf P}_0), \nonumber\\ Y_0
=Y({\bf r}, {\bf p}, t_0),\\ {\bf P}_0 = {\bf P}({\bf r}, {\bf p},
{\bf t_0}).\nonumber
\end{eqnarray}

Here, $f_0(y,{\bf p})$ is an injected distribution where all the
injected electrons have a characteristic energy $\varepsilon_0$.
The quasi-particles, which appear as the result of a collision,
are described by the action of the incoming part $\hat{\nu}$ of
the collision operator $\hat{J}$ on $f_0(y, {\bf p })$. In other
words, $\hat{\nu} f_0$ correspond to a particle that has been
scattered once, while $\hat{\nu}\hat{\nu} f_0$ describes a
subsequent second scattering event, and so forth. Correspondingly,
the typical energies of these states differ: $f_0$ is
characterized by the energy $\varepsilon_0$, for $\hat{\nu} f_0$
the characteristic energy will be $\varepsilon_{1} \sim
\varepsilon_{0} \cdot \Delta$ and for $\hat{\nu} \hat{\nu} f_{0}$
it will be given by $\varepsilon_{2} \sim \varepsilon_{0} \cdot
\Delta^2$ etc. Consequently we find that $v^{-1}\nu \hat{\nu}
f_{0} \sim l^{-1}(\varepsilon_{1}) \hat{\nu} f_{0}$ and that
$\hat{\nu} v^{-1} \nu f_{0} \sim l^{-1}(\varepsilon_{0}) \hat{\nu}
f_{0}$. Therefore, as a main approximation in the parameters
$l(\varepsilon_0)/l(\varepsilon_1)$, $L/l(\varepsilon_1) \ll 1$,
it is possible to neglect all terms in the series of
Eq.~(\ref{e-mocriad}) that include $\hat{\nu}$ or $\nu$ to an
order higher than one, where $\nu$ is placed to the left of
$\hat{\nu}$. After making this approximation, the summation of the
remaining terms can be carried out easily. Thus, the main idea
consists in a partial summation of the infinite series to a first
or subsequent approximation, as required in order to reach
sufficient accuracy for an adequate description of the experiment.

For a stationary beam in a magnetic field, a Green's function of
the "MOC-1" (i.e. the solution of Eq.~(\ref{e-kin}) for
$f_0(y,{\bf p}) = \delta(y) \delta({\bf p} - {\bf p}_0)$, that
takes into account the first collision exactly) has the following
form:

\begin{equation}\label{e-moca}
g({\bf r}, {\bf p}) = g_b + \sum \nu_{\bf P P'}\frac{V_{0
x}}{|{\bf V_1} \times {\bf V'}|} \exp{\{\nu(\varepsilon_0)(t_0 -
t')\}} \ .
\end{equation}
Here $g_b$ is the distribution of the ballistic part of the
electrons, i.e., the solution of Eq.~(\ref{e-kin}) in the absence
of the integral operator $\hat{\nu}$. $t'$ is the time of
collision and $t_0$ is the starting time. ${\bf P}' = m {\bf V'}$
and ${\bf P_1} = m {\bf V_1}$ are the momenta before and after
collision, respectively, while ${\bf V}_0 = {\bf P_0}/m$. The
values of $t$, $t_0$, ${\bf R}({\bf r}, {\bf p}, t')$, ${\bf P' =
{\bf P}(0,{\bf p}_0, t_0 - t' )}$, ${\bf P_1} = {\bf P}({\bf r},
{\bf p}, t')$, and ${\bf p}$ are determined through the equations
of motion by the initial momentum ${\bf p}_0$ and location ${\bf
r}$. The equations of motion are reduced to ${\bf R}({\bf r}, {\bf
p}, t') = {\bf R}(0, {\bf p}_0, t_0 - t')$. The presence of walls
in the experimental setup, see Fig.~\ref{g-exp}, excludes
solutions with a full cyclotron orbit.

Note that the MOC approximation is applicable in principle under
conditions, where a pre-temperature stage occurs for the
non-equilibrium quasi-particles. In other words, it holds when the
following hierarchy of mean-free paths is given:

\begin{equation}\label{e-lierarh}
  l(\varepsilon_0) \ll l(\varepsilon_1) \ll l(\varepsilon_2) \ll l(\varepsilon_3) \ll
  ... \lesssim l(T),
\end{equation}
where $\varepsilon_i$ is the energy after the $i$-th collision and
$ l(\varepsilon_0) \lesssim L  \ll l(T)$; for $L\ll
l(\varepsilon_0)$ the inequality corresponds to the
quasi-ballistic regime, where perturbation theory is applicable.
If $L$ corresponds to the $n$-th segment of this hierarchy, i.e.
$l(\varepsilon_n) < L < l(\varepsilon_{n+1})$, it is necessary to
construct a "MOC-$n$" approximation by treating the first $n$
collisions exactly; which leads to relative corrections of the
order of $\Delta^2$. The approach breaks down when the
quasi-particle energy is comparable to $T$. As an example, the
exact result for the first two collisions ("MOC-2"), for the
spatially homogeneous case and $f _ {0} \propto \delta
(\varepsilon - \varepsilon _ {0}) $, reads:

\begin{multline}\label{e-moca2}
  f = \Big\{ \exp(-\nu_{0} t)+(\nu_{0}-\nu)^{-1} \Big(\exp(-\frac{\nu
  t}{1-\nu/\nu_{0}})\\- \exp(-\nu_{0} t) \Big) \hat{\nu} +
    \nu^{-1}_{0} \hat{\nu} \Big[\nu^{-1} (1 - \exp(-\nu t)) \\+
  (\nu_{0} - \nu)^{-1}\big(
   \exp(-\nu_{0} t) - \exp(\frac{\nu t}{1-\nu/\nu_{0}})\big) \Big]\hat{\nu}
   \Big\}f_{0},\hspace{1mm}\\
   \nu_{0}=\nu(\varepsilon_{0}).
\end{multline}
This formula is valid for times $t < l(\varepsilon_2)/v_F$. Thus,
just a few steps lead to a high accuracy. It turns out, that for
the experiments discussed below, just one step is enough to
produce satisfactory results.

\section{Experiment}
In order to investigate the electron-electron scattering
properties in a 2DEG, experiments are needed which allow to
distinguish between scattered and non-scattered particles, and
reveal the details of scattered signal. Therefore, we used the
experimental setup of the electron spectroscopy method proposed in
Ref.~\cite{epl2001} (see \ Fig.~\ref{g-exp}). The sample is based
on a conventional Si-modulation doped Ga-As(Al,Ga)As
heterojunction with a carrier density of $n \approx 2.8 \cdot
10^{11}$ cm$^{-2}$ and an impurity mean-free path of $l_{im}
\approx 20$ $\mu$m. The scheme of the experiment is shown in
Fig.~\ref{g-exp}. Schottky gates define quantum point-contacts in
the 2DEG and are adjusted such that the injector and detector
quantum point-contact (QPC) are carrying only one conducting mode.
The distance between the injector and detector QPC is $L=4$
$\mu$m. Ohmic contacts are indicated by crossed squares in
Fig.~\ref{g-exp}. The sample was cooled down to $T=0.2$ K. An
ac-modulated dc-bias voltage applied between ohmic contacts 1 and
2 was used to inject the electrons into the 2DEG area between
injector and detector. The amplitude of the ac modulation $\delta
V_i$ was much smaller than the dc component, $\delta V_i \ll V_i$.
The ac-component of the detector signal, i.e.\ the voltage that
occurs between contacts 3 and 4 ($V_d = V_{34}$) is measured with
a lock-in technique. With this technique only the signal of
electrons with energy $V_i$ is measured, while the whole e-beam
consists of electrons with energies in the range of $0\dots V_i$.
(Note that, throughout this paper the electron charge $e$ is set
equal to one $e=1$.)
\begin{figure}
\includegraphics[height=6 cm,width=8cm]{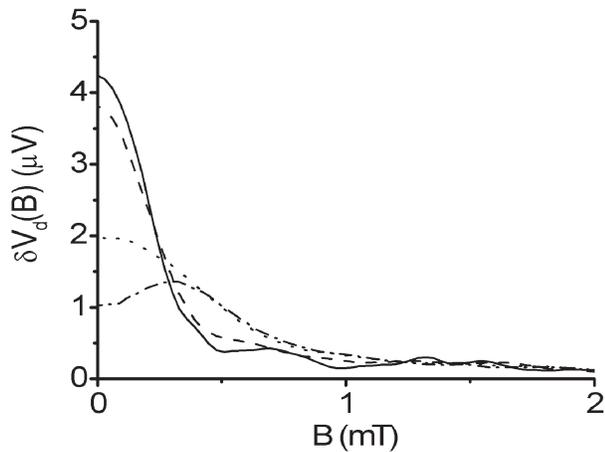}
\caption{The experimentally measured signal on the detector as a
function of magnetic field for different energies of injection:
(solid) -- $0.8$ mV, (dash) -- $1.6$ mV, (dot) -- $2.6$ mV,
(dash-dot) -- $3.4$ mV.} \label{g-expB}
\end{figure}

The magnetic field dependence of the detector signal $\delta
V_{d}(B, V_{i})$ has been studied for a wide range of energies
$V_i$ $=$ $0\dots 9$ mV. An example is given in Fig.~\ref{g-expB}
for different injection energies $V_i$. This figure display the
evolution of an mono-energetic electron beam with increasing
injection energy$ V_i$, which reflects the energy-dependent
electron-electron scattering.  From the half-hight width of low
energy curve (1) ($V_i=01.$ mV) it is possible to extract the
opening angle of the collimated electron beam: $\Phi \approx
12^\circ$, which is identical for injector and detector when the
QPCs are equally adjusted to one conducting mode ($G_{\rm
QPC}=2/h$).

The contribution of scattered particles to the detector signal
$\delta V^{s}_{d}$ can be easily extracted following the approach
of Ref.~\cite{epl2001}:

\begin{equation}\label{e-balextr}
    \delta V^{s}_{d}(B, V_{i}) = \delta V_{d}(B, V_{i}) - \exp \Big( - \frac{2 r_c}{l(V_{i})}
    \arcsin \frac{L}{2 r_c} \Big)\delta V^{0}_{d}(B) ,
\end{equation}
here, $l(V_i) = 4 \pi \hbar v_F (\varepsilon_{F}/V_{i})^{2}/
\varepsilon_{F} (\ln \varepsilon_{F}/V_{i}+\ln 2
q_{TF}/k_{F}+1/2)$ is the energy dependent e-e scattering length
according to Ref.~\cite{quinn}, where $q_{TF}$ is the Thomas-Fermi
screening wave vector; $r_c$ is the cyclotron radius; $v_F$ is the
Fermi-velocity. The second term on the right-hand side of this
expression corresponds to $g_b$ defined in Eq.~(\ref{e-moca}),
$\delta V^{0}_{d}(B)$, represents the signal that would be
observed in the absence of scattering. A good approximation for
this value can be obtained from the signal measured for the lowest
excitation voltage $\delta V_{d}(B, V_{i})$ at $V_{i} = 0.1$ mV.
In this cases, $L/l \sim 10^2$ and electron-electron collisions
can be ignored. Note that for injection energies $V_i>3.5$ mV
$\approx \varepsilon_F/3$ the above equation for $l(V_i)$ yields a
large inaccuracy. However, this can be ignored because $l(V_i)$
becomes so small that the subtraction of the ballistic
contribution to the signal leads only to corrections of less than
one percent \cite{epl2001}.
\begin{figure}
\includegraphics[height=7 cm,width=8 cm]{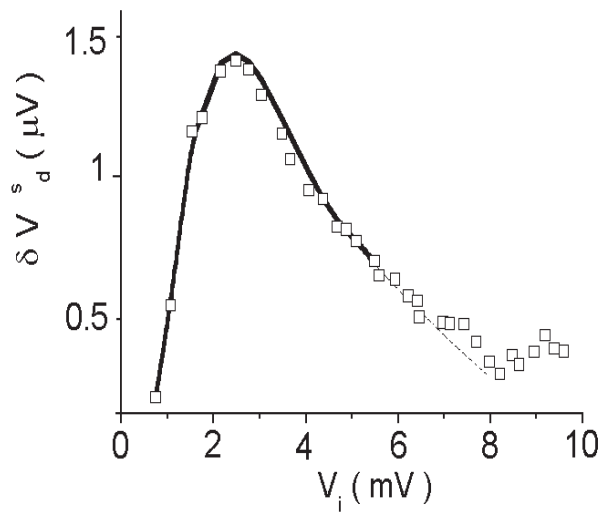}
\caption{The maximum of scattered particles signal as a function
of injection energy. Empty square markers correspond to the
experimental values. The thick line represents the theory. The
dashed line represents theoretical calculations outside the range
of applicability of the approximation used here.}\label{g-vsmx}
\end{figure}

\section{Discussion}
In Fig.~\ref{g-vsmx} the maximum value of the magnetic field
dependent signal due to the scattered electrons, $\delta
V^s_{max}(V_i) = \max_{B} \delta V_d^s(B,V_i)$ is presented as a
function of the injection energy ($V_i$). This dependence shows a
distinct non-monotonic behavior, i.e. a rapid increase at small
$V_{i}$, a distinct maximum at $V_i \approx 2$ mV and an abrupt
decrease for larger injection energies up to $V_i \sim 8$ mV,
followed by a weak increase.

The presence of a maximum and the subsequent decrease in the
energy dependence of $\delta V^{s}_{max}(V_i)$ can not be
explained in the context of a simple perturbation theory applied
to the collision integral, because in the regime where the maximum
is observed the electron-electron mean-free path is already
smaller than the distance between the injector and detector QPC.
Perturbation theory can only be applied to describe the increase
of the signal for small energies $V_i$. The presence of a strong
decrease appears at first sight to be unexpected, because in the
multi-collision regime that is entered upon increasing $V_i$, a
decrease of the ac-signal from scattered particles is not
anticipated. In this multi-collision regime the ballistic beam is
already absent but the motion of the electrons can be represented
by a hydrodynamic drift.

However, the behavior shown in Fig.~\ref{g-vsmx} can be explained
in the context of the modified one-collisional approximation
described above by taking into account the specific properties of
e-e scattering in two-dimensions \cite{gfnt97}. A calculation
shows, that by taking into account the first collision exactly,
Eq.~(\ref{e-moca}) leads to the following expression for the
detector signal due to scattered particles

\begin{equation}\label{e-specform}
\delta V^{s}_{d} \propto \nu K(\psi) G(\psi) \exp \big(
-\frac{L}{4l} \frac{\psi}{\sin \psi/2}\big) , \  \psi = 2 \arcsin
\frac{L}{2 r_c},
\end{equation}
where $\psi$ is the scattering angle and $G(\psi)=
\int_{-\Phi/2}^{\Phi/2} d \varphi g(\psi - \varphi)$ is the
angular distribution function of e-e scattering, $g(\psi)=
m\nu^{-1}\int d \varepsilon \, \nu_{\varepsilon
\varepsilon_0}(\psi)$, averaged over an interval $\Phi$; the
weighting coefficient $K$ is $K \approx 1/\Phi$ for $\psi \lesssim
\Phi$ and $K \approx 1 / \sin \psi$ for $\Phi<\psi<\pi/2$. (It is
sufficient to consider only the case where the cyclotron radius
$r_c$ is smaller than $L$.) It should be noted that
Eq.~(\ref{e-specform}) is valid for all ratios of $\Phi$ and the
angular width of $g(\psi)$. The expression $L\psi/4 \sin(\psi/2)$
in the exponent of Eq.~(\ref{e-specform}) represents the length of
the electron trajectory from the injector to the point $O$ where
the scattering takes place (see \ Fig.~\ref{g-exp}). From
Eq.~(\ref{e-specform}) it follows, that the signal due to
scattered particles is directly proportional, on the one hand, to
the e-e scattering probability, or the frequency of the collisions
$\nu$ and, on the other hand, it is proportional to the
probability of an electron to reach the scattering point $O$
ballistically. As has been shown earlier in \cite{epl2001}, the
pre-exponential factor increases with increasing $V_i$ at low
injection energies (mainly due to the increase of the collision
frequency $\nu \propto V_i^2$).

The existence of a maximum in Fig.~\ref{g-vsmx} can be attributed
to a competition between an increase of the scattering probability
(the pre-exponential factor) and a decrease of ballistic transport
into the neighborhood of the scattering point $O$ (the exponent)
with $V_i$. This explanation is supported by the excellent
quantitative agreement between the experimental data and a
theoretical fit based on Eq.~(\ref{e-specform}), where the angle
$\psi$ is replaced by the experimental value $\psi_{max}$, see
Fig.~\ref{g-vsmx}.

The good fit also justifies the use of the one-collision MOC
approximations indicating that higher orders collisions can be
neglected. For small energies, the pre-exponential factor
determines the behaviour of the signal due to scattered particles,
which also can be described by standard perturbation theory
applied to the collisions integral. Around the maximum the
probability for secondary collisions is still small and the signal
is still defined by the scattering that takes place in the
neighborhood of the point $O$ - however, the probability for a
first collision on the way between the injector and the detector
becomes larger. This underlines the decrease of the signal with
increasing $V_i$. As mentioned above this regime can not be
described by ordinary perturbation theory, requing the MOC
approximation. At this point we would like to note that the
non-monotonic behavior of the ac-current component which has been
discussed above can not be explained by nonlinearities due to
heating effects which are expected to appear at approximately the
same values of $V_i$ in dc-measurements \cite{gprb2000}. It can be
shown, that such an effect gives only a small correction of the
order of $T(V_i) / V_i$ to the ac-component of the signal, where
$T(V_i)$ is the temperature due to the heating effect.

Finally, we may only speculate about the origin of the weak
increase of the detector signal which is observed for the highest
injection energies $V_i$. As the injection energy approaches
$\varepsilon_F$, the potential drop across the injector QPC
becomes so large that the second subband of the QPC may become
populated. This enhances the injector conductance, and,
consequently, the detected signal. For these energies
Eq.~(\ref{e-specform}) is no longer valid (see Ref.
\cite{epl2001}). A second reason could be transition to the
two-collision (or multi-collision) regime as the mean-free path is
less then $10^{-1} L$ at these energies.

\section{Conclusion}

In summary, we have shown in this paper that it is possible to
successfully investigate the evolution of a high-energy
non-equilibrium quasi-particles distribution with the help of the
"modified one-collision approximation". This method was used to
explain novel measurements of the non-monotonic energy dependence
of the signal of scattered electrons in a 2D system. We have
obtained an excellent quantitative agreement between the
experimental data and a theoretical calculation.

In conclusion we would like to emphasize that the MOC
approximation may be of interest for a wide range of problems
independent of the nature of the scattering particles; the only
restriction is that the scattering probability should strongly
decrease after one or several collisions. Along with the case of
electron-electron scattering, this is a typical situation for the
majority of inelastic scattering processes, such as
electron-phonon, phonon-phonon etc. Let us discuss as an example,
electron-phonon scattering in the case when the electron energy is
high enough $\varepsilon \gg T$, but is low in comparison with the
Debye temperature (the electron energy is measured from the Fermi
level). In this case a phonon with an energy of the order of
$\varepsilon$ will be emitted as a result of the electron-phonon
scattering event and the corresponding mean-free path is $l
\propto \varepsilon^{-3}$. In contrast to electron-electron
scattering the energy is redistributed here between two states
according to the momentum and energy conservation laws, ${\bf p_0}
= {\bf p_1} + {\bf q} $ and $\varepsilon_0 = \varepsilon_1 +
\omega$. Here ${\bf p_0}$ and $\varepsilon_1$ are the
quasimomentum and energy before the collision, and after the
collision they are ${\bf p_1}$, and $\varepsilon_1$, respectively;
${\bf q}$ is quasimomentum of the emitted phonon and $\omega$ its
energy. Since, on average, the energy is redistributed equally
between the scattering product $\omega \sim \varepsilon_1$, we
also have here the small parameter $l(\varepsilon_0) /
l(\varepsilon_1) \sim 2^{-3} \simeq 0.1$. The case of
three-particles high-energy phonon-phonon scattering processes
differs only by the energy dependence of the mean-free path.

Furthermore, the inelasticity, on which the approach developed in
this paper is based, is likely to be relevant not only to the
energy redistribution but also to some other quantities; this
includes, for example, the spin in cases where the probability of
scattering decreases significantly after a spin-flip scattering
event. This case may be realized for electron scattering from
spin-polarized impurities (there are no spin subbands in this
case) and if the exchange interaction depends significantly on the
mutual orientation of the electron spin and the impurity spin. Let
us neglect the interaction between electrons of one of the spin
directions (spin-up, for example) and the impurities. Then the
spin relaxation process is completed just after the first
spin-flip scattering of the spin-down electrons. Thus, it is
straightforward  to describe this process in the framework of the
modified one-collision approximation.

The method we proposed in this work can be very effective also for
point-contact spectroscopy \cite{yan,kul}. The reason is the
following. The current-voltage characteristic of the electric
circuit containing a microconstriction is determined by the
back-scattering due to inelastic electron scattering processes
just in the region of the microconstriction. If
$l(\varepsilon)\lesssim L$ (L is the length of the constriction)
the current-voltage characteristic is determined only by one, or
few, first collisions, the probability of which is not small and,
thus, it can not be taken into account using a perturbation theory
(as it was done in the study described in Ref.\cite{kul} for the
case when $l(\varepsilon)\gg L$). The same also holds for the
nonlinear thermoelectric phenomenon in point-contacts \cite{ed}.
The influence of the first collisions on the point-contact
spectrum may be calculated exactly using equations of the general
form  of Eq.(6) because in point-contact spectroscopy the excess
energy of the electron is high, $\varepsilon \gg T$.

\section*{Acknowledgement}
The research described in this publication was made possible in
part by Award No.UP2-2430-KH-02 of the U.S. Civilian Research \&
Development Foundation for the Independent States of the Former
Soviet Union (CRDF)(R.N.G, A.N.K, A.I.K., A.V.Y., E.N.B., and
U.L.), by the US Department of Energy Grant FG05-86ER-45234
(E.N.B. and U.L.), and by the Deutsche Forschungsgemeinschaft SFB
410 (H.B. and L.M.).


\begin{thebibliography}{99}

\bibitem{msst} L.W. Molenkamp, M.J. Brugmans, H. van Houten and C.T. Foxon,
    Semicond. Sci.  Technol. {\bf 7}, B228, (1992).
\bibitem{juapp}Th.Sch\"apers, M.Kr\"uger, J.Appenzeller, A.F\"orster, B.Lengeler,
    and H.L\"uth, Appl. Phys. Lett. {\bf 66}, 3603 (1995).
\bibitem{cumming} D.R.S. Cumming and J.H. Davis, Appl. Phys. Lett.
{\bf 69}, 3363 (1996).

\bibitem{epl2001}
    A.V. Yanovsky, H. Predel, H. Buhmann, R.N. Gurzhi, A.N. Kalinenko,
     A.I. Kopeliovich, L.W. Molenkamp, Europhys.
     Lett., {\bf 56}, 709, (2001).
\bibitem{chaplik} A.V. Chaplik, Zh. Eksp. Teor. Fiz.,  {\bf 60},1845
(1971).

\bibitem{quinn} G.F. Giuliani, and J.J. Quinn, Phys.Rev. {\bf B}, {\bf 26}, 4421 (1982).

\bibitem{gfnt97}   R.N. Gurzhi, A.N. Kalinenko, A.I. Kopeliovich,
    Fiz. Nizk. Temp., {\bf 23}, 58, (1997) [Low Temp. Phys. {\bf
    23}, 44, (1997)]; Surface Science. {\bf 361/362}, 497,
  (1995).

\bibitem{gprl}R.N. Gurzhi, A.N. Kalinenko, A.I. Kopeliovich,
    Phys. Rev. Lett. , {\bf 74}, 3872, (1995); Phys. Rev. B, {\bf 52}, 4744 (1995).
\bibitem{gprb2000}H. Predel, H. Buhmann, R.N. Gurzhi, A.N. Kalinenko, A.I. Kopeliovich,
     L.W. Molenkamp, and A.V. Yanovsky, Physical Review {\bf B}, {\bf 62}, 2057, (2000).
\bibitem{gphyse98} H. Buhmann,  R.N. Gurzhi,  A.N. Kalinenko,  A.I.
    Kopeliovich,  L.W. Molenkamp, H. Predel,  A.V. Yanovsky, Physica
    E, {\bf 6}, 310, (2000).
\bibitem{yan} I.K. Yanson, Zh.Eksp.Theor.Fiz. {\bf 66}, 1035
(1974)[Sov.Phys.JETP {\bf 39}, 506 (1074)].
\bibitem{kul} I.O. Kulik, A.N. Omelyanchuk, and R.I.Shekhter, Fiz. Nizk. Temp., {\bf 3}, 1543, (1977) [Sov. J. Low
    Temp. Phys. {\bf 3} 789, (1977)].
\bibitem{ed} E.N. Bogachek, I.O. Kulik, and A.G.Shkorbatov, Fiz. Nizk. Temp., {\bf 11}, 1189, (1985) [Sov. J. Low
    Temp. Phys. {\bf 11} 656, (1985)].

\end{thebibliography}
\end{document}